\documentclass[reprint,amsmath,amssymb,twocolumn]{aastex631}
\usepackage{aas_macros}
\usepackage{graphicx} % include figure files
\usepackage{hyperref} % add hypertext capabilities
\usepackage{amsmath}

\newcommand{\rms}{r_\mathrm{ms}}
\newcommand{\rph}{r_\mathrm{ph}}
\newcommand{\rp}{r_\mathrm{p}}
\newcommand{\arccosh}{\cosh^{-1}}

\begin{document}

\title{Shadow Geometry of Kerr Naked Singularities}

\author{Bao Nguyen}
\email{bnguyen20@arizona.edu}
\affiliation{Department of Astronomy, University of Arizona, 933 North Cherry Avenue, Tucson, AZ 85721, USA}
\affiliation{Department of Physics, University of Arizona, 1118 East Fourth Street, Tucson, AZ 85721, USA}
\affiliation{Department of Mathematics, University of Arizona, 617 North Santa Rita Avenue, Tucson, AZ 85721, USA}

\author{Pierre Christian}
\affiliation{Physics Department, Fairfield University, 1073 N Benson Rd, Fairfield, CT 06824}

\author{Chi-kwan Chan}
\affiliation{Department of Astronomy, University of Arizona, 933 North Cherry Avenue, Tucson, AZ 85721, USA}
\affiliation{Data Science Institute, University of Arizona, 1230 N. Cherry Ave., Tucson, AZ 85721}
\affiliation{Program in Applied Mathematics, University of Arizona, 617 N. Santa Rita, Tucson, AZ 85721}

\begin{abstract}
Direct imaging of supermassive black holes (SMBHs) at event horizon scale resolutions, as recently done by the Event Horizon Telescope, allows for testing alternative models to SMBHs such as Kerr naked singularities (KNSs). We demonstrate that the KNS shadow can be closed, open, or vanishing, depending on the spins and observational inclination angles. We study the critical parameters where the KNS shadow opens a gap, a distinctive phenomenon that does not happen with the black hole shadow. We show that the KNS shadow can only be closed for dimensionless spin $a \lesssim 1.18$ and vanishing for $a \gtrsim 1.18$ for certain ranges of inclination angles. We further analyze the effective angular momentum of photon orbits to demonstrate the fundamental connections between light geodesics and the KNS shadow geometry. We also perform numerical general relativistic ray tracing calculations, which reproduce the analytical topological change in the KNS shadow and illustrate other observational features within the shadow due to the lack of an event horizon. By comparing with black hole shadow observations, the topological change in the shadow of KNSs can be used to test the cosmic censorship hypothesis and KNSs as alternative models to SMBHs.
\end{abstract}


% \renewcommand{\section}{\paragraph}

%==============================================================================
\section{Introduction}
\label{sec:intro}

The Event Horizon Telescope (EHT) recently resolved the supermassive black holes (SMBHs) at the center of Messier 87 (M87*) and the Milky Way (Sagittarius A*, or Sgr~A*) at the event horizon scale \citep{2019ApJ...875L...1E, 2019ApJ...875L...2E, 2019ApJ...875L...3E, 2019ApJ...875L...4E, 2019ApJ...875L...5E, 2019ApJ...875L...6E, SgrA1, SgrA2, SgrA3, SgrA4, SgrA5, SgrA6}. The reconstructed black hole images show a bright asymmetric ring surrounding an interior brightness depression. These observations were used to test modified gravity theories and alternative models of galactic central compact objects by placing constraints on deviations from the Kerr metric using the size and shape of the SMBH shadows \citep{2020PhRvL.125n1104P}.

If M87* and Sgr~A* are Kerr black holes (KBHs), then they possess unstable spherical photon orbits, which separate capture orbits from scattering orbits, thus casting a shadow~\citep{2010ApJ...718..446J}. This signature of the KBH is demonstrated by the radius and circularity of the bright photon rings shown on the images of M87* and Sgr~A*, which are mostly independent from the accretion profiles and instead largely dependent on the spacetime surrounding the SMBHs \citep{2019ApJ...875L...1E, 2019ApJ...875L...5E, 2019ApJ...875L...6E, SgrA1, SgrA5, SgrA6, 2000ApJ...528L..13F, 2011PhRvD..83l4015J, 2019ApJ...885L..33N, 2019PhRvD.100b4018G, 2021arXiv211101752Y, 2021ApJ...920..155B, 2022MNRAS.513.1229K}. While a KBH is a sufficient condition for a shadow image, the converse is not necessarily true because other theoretical objects, such as naked singularities with certain physical parameters, can also project shadow-like regions without event horizons and photon spheres \citep{2019MNRAS.482...52S, 2021PhRvD.103b4015D}. Thus, the EHT concludes that the possibility of Sgr~A* being a naked singularity cannot be ruled out based on the shadow-based metric tests~\citep{SgrA6}.

The weak cosmic censorship conjecture demands that all singularities from gravitational collapse are expected to be hidden by event horizons. Evidence for violations of this conjecture in nature will thereby have important implications for fundamental physics \citep{1969NCimR...1..252P, 1997gr.qc....10068W}. There are multiple valid naked singularity spacetime solutions to the Einstein field equations with feasible formation mechanisms from gravitational collapse \citep{1968PhRvL..20..878J, 1984CMaPh..93..171C, 1991PhRvL..66..994S, 2011CQGra..28w5018J, 2017PhRvL.118r1101C}. Among singularity spacetimes without a horizon, Kerr naked singularities (KNSs) have emerged as a potential candidate to model SMBH images. According to the weak cosmic censorship conjecture, a black hole with mass M and angular momentum $J$ must satisfy the Kerr bound $|a| \leq 1$ to keep the gravitational singularity hidden by an event horizon, where $a = J/M^2$ is a dimensionless rotational parameter \citep{1969NCimR...1..252P}. There are theoretical arguments that the Kerr bound can be violated in string theory and observations of KNSs can provide direct experimental evidence for string theory \citep{2009PhLB..672..299G}.

The linear instability of KNSs in general relativity (GR), as demonstrated by \cite{2008CQGra..25s5010C, 2008CQGra..25x5012D, 2018PhLB..780..410N}, motivates theoretical analysis of KNS shadows as an observational framework to test deviation from GR. If the EHT or future horizon-scale imaging experiments detect signatures of KNSs, one or more of the assumptions underlying GR must be violated in nature. Studies of KNS shadows have similar implications as previous research on the observational signatures of deviation from general relativity predictions. For example, \cite{2004PhRvD..69l4022C, PhysRevD.83.124015, 2014PhRvD..90h4009R} considered observational signatures of metrics which are not valid solutions of the vacuum equations in GR. \cite{Jusufi:2021lei} used observations of stellar motion in the galactic center and Sgr~A* shadow imaged by the EHT to propose wormhole solutions, many of which are unstable and/or require stress-energy that violates the energy conditions, as candidates of Sgr~A*. Nevertheless, they still provide rigorous tests of GR with black hole observations. Similarly, though KNSs are disfavored models for compact objects, if they can be identified in nature, they would provide a strong observational evidence that GR demands modifications in the strong-field regime.

There have been many theoretical works on KNSs. The conversion from KNSs to black holes due to the instability of their spacetime was demonstrated to be slow enough that primordial KNSs may exist for a certain range of cosmological redshifts \citep{2011CQGra..28o5017S}. Classification of KNS spacetimes based on different characteristics of the spherical photon orbits was done in \cite{2018EPJC...78..879C}. There are studies on the topological properties of shadows in Kerr spacetime, including both KBHs and KNSs, which discusses how spins and observational inclination angles of Kerr compact objects relate to observable quantities such as the arc length, angle, curvature radius, and defined distortion parameter of shadows \citep{2009PhRvD..80b4042H, 2019PhRvD..99d1303W}. \cite{Tavlayan:2023vbv} argues that observers with face-on inclinations cannot distinguish between a KNS shadow and a KBH shadow for a certain range of spins slightly above 1. The interaction between null orbits and timelike orbits and its observational consequence was investigated in \citep{2018EPJC...78..879C}. In addition, the effects of a repulsive gravitational effect near the singularity to the accretion process onto KNSs were considered in \citep{2009PhRvD..80j4023B}. Some optical features of KNSs such as the accretion disks and the spectral lines were studied \citep{2010CQGra..27u5017S, 2013JCAP...04..005S}.

KNSs have been considered as black hole mimicker candidates for recent shadow observations from the EHT. The shadow of M87* is measured to be very circular, with deviations from circularity of about $10\%$ or less in terms of the root-mean-square distance from the average radius of the shadow \citep{2019ApJ...875L...1E}. Meanwhile, \citet{2009PhRvD..79d3002B} suggests that the analytical apparent shape of KNSs is elliptical or crescent-like, depending on the observational inclination angle and spin, which prompted the EHT to rule out the possibility that M87* is a KNS~\citep{2019ApJ...875L...6E}. However, \citet{2019PhRvD.100d4057B} refutes this claim by analytically demonstrating that the inferred size and circularity of the observed shadow of M87* can still be produced by KNSs with certain quantum effects. Currently, the EHT has not placed constraints on the circularity of the shadow of Sgr~A* because of substantial observational uncertainties \citep{SgrA6}. Despite so, future generations of EHT with added telescopes are expected to measure the circularity of the shadow of Sgr~A* and conduct further tests for deviations from the Kerr metric \citep{SgrA6}, providing another prospect in testing the possibility that Sgr~A* is a KNS. Furthermore, the close proximity of Sgr~A* allows extremely accurate inference of its mass and distance from Earth, which is important in relating the angular size of its photon ring with the properties of its shadow, allowing highly sensitive tests of gravity in the strong-field regime \citep{2011JPhCS.283a2030P, 2016PhRvL.116c1101J}. Recent studies have taken advantage of the precise measurements of the mass and distance of Sgr~A* to test a variety of black hole models, modified theories of gravity, and physically motivated alternative candidates of SMBHs including naked singularities \citep{2022arXiv220507787V, 2022arXiv220605878K, 2022arXiv220602488G}. These analysis have provided strong preliminary constraints on the different possibilities of the nature of Sgr~A* and suggested promising future directions for further gravity tests with Sgr~A* observations \citep{SgrA1, SgrA6}.

In this paper, we perform a systematic study of the projection of the unstable spherical photon orbits surrounding KNSs at infinity, which we define to be the ``shadow''. In Section~\ref{sec:shadow}, we analytically calculate the KNS shadow by separating the Hamilton-Jacobi equation and analyzing the radial effective potential. We demonstrate that the KNS shadow can be closed, open, or vanishing, depending on the spins and observational inclination angles, and present the critical parameters where the shadow changes its topology. Furthermore, we analyze the effective angular momentum of photon orbits to demonstrate the fundamental connections between light geodesics and the KNS shadow geometry. In Section~\ref{sec:numerical}, we describe our numerical setup where we integrate null geodesics backward in time in Cartesian Kerr-Schild spacetime. We then discuss the observational signatures of KNSs based on our numerical ray tracing calculations. We present deflection angles of the null geodesics to illustrate how the KNS shadow alters due to different spins and observational inclination angles and compare between our numerical results and analytical predictions in section~\ref{sec:shadow}. In Section~\ref{sec:discussions}, we discuss the implications of our results in shadow-based metric tests and constraints to KNSs as alternative models to SMBHs from horizon-scale imaging like the EHT.

%==============================================================================
\vspace{-0.5cm}
\section{Analytical Shadow of KNS}
\label{sec:shadow}

\subsection{Unstable Spherical Photon Orbits Around KNS}
\label{sec2a}

In our study, we use geometric units $c = G = M = 1$. The Kerr metric is an axially symmetric, stationary vacuum solution to the Einstein field equations, which describes uncharged rotating compact objects, whose mathematical formulation is derived in \cite{1963PhRvL..11..237K}. The line element of the Kerr metric in Boyer-Lindquist coordinates ($t, r, \theta, \phi$) is \citep{1967JMP.....8..265B}: \vspace{-0.5cm}
\begin{align}\label{eq1}
    ds^2 = \left( \frac{2Mr}{\Sigma} - 1 \right) dt^2 &- \frac{4Mar}{\Sigma} \sin^2 \theta \, dt \, d\phi \nonumber \\
    &+ \frac{\Sigma}{\Delta} + \Sigma \, d\theta^2 + \frac{\beta}{\Sigma} \sin^2 \theta \, d\phi^2
\end{align}
where
\begin{subequations}
\begin{align}
    \Sigma &= r^2 + a^2 \cos \theta, \label{eq2a} \\
    \Delta &= r^2 + a^2 - 2Mr, \label{eq2b} \\
    \beta &= (r^2 + a^2)^2 - \Delta \, a^2 \sin^2 \theta. \label{eq2c}
\end{align}
\end{subequations}
Here, $a = J/M^2$ is a dimensionless spin for a Kerr compact object with mass $M$ and angular momentum $J$. For $|a| < 1$, the metric describes a KBH; and for $|a| > 1$, the metric describes a KNS. We further consider only $a > 0$ without loss of generality.

Because the Kerr metric is stationary and axially symmetric, there are two Killing vectors $K^{\mu} = (1, 0, 0, 0)$ and $R^{\mu} = (0, 0, 0, 1)$ representing the time and azimuthal translation. Conserved quantities can be derived from the symmetries of the physical laws. By projecting the Killing vectors along the covariant four-momentum vector $p_{\mu}$, we obtain energy $E$ and angular momentum in the $\phi$ direction $L_z$:
\begin{subequations}
\begin{align}
    E &= -K^{\mu} p_{\mu} = -p_{t}, \label{eq3a} \\
    L_z &= R^{\mu} p_{\mu} = p_{\phi}. \label{eq3b}
\end{align}
\end{subequations}

To analyze the spherical photon orbits and the shadow of KNSs, we employ the Hamilton-Jacobi equation:
\begin{align} \label{eq4}
    \frac{\partial S}{\partial \lambda} = \frac{1}{2} g^{\mu \nu} \frac{\partial S}{\partial x^{\mu}} \frac{\partial S}{\partial x^{\nu}}
\end{align}
Here, $S$ is the action as a function of the affine parameter $\lambda$ and coordinates $x^{\mu}$. The solution to equation~(\ref{eq4}) can be separated into different components that only depend on each of the Boyer-Lindquist coordinates \citep{1968PhRv..174.1559C}:
\begin{align} \label{eq5}
    S = -Et + S_r(r) + S_{\theta}(\theta) + L_z \phi
\end{align}

Equations~(\ref{eq4}) and~(\ref{eq5}) yield the following equations of motion for null geodesics for each of the coordinates \citep{1979GReGr..10..659S}:
\begin{subequations}
\begin{align}
    \Delta \Sigma \dot{t} &= aE - 2MaL_zr \label{eq6a} \\
    \Sigma^2 \dot{r}^2 &= E^2 R(r) \label{eq6b} \\
    \Sigma^2 \dot{\theta}^2 &= C + (a^2 E^2 - L_z^2 \csc^2 \theta) \cos^2 \theta \label{eq6c} \\
    \Delta \Sigma \dot{\phi} &= 2MaEr + (\Sigma - 2Mr)L_z \csc^2 \theta \label{eq6d},
\end{align}
\end{subequations}
where the overdots represent derivatives with respect to the affine parameter $\lambda$ along the geodesics. The radial equation of motion is written in terms of the radial effective potential $R(r)$, which is
of major interest in this study \citep{stewart}:
\begin{align}\label{eq7}
    R(r) = r^4 &+ r^2 (a^2 - \Phi^2 - Q) \nonumber\\
    &+ 2Mr[(a - \Phi)^2 + Q] - a^2 Q.
\end{align}

Here, we define impact parameters $\Phi = L_z/E$ and $Q = C/E^2$, where $C$ is the Carter's constant, a third conserved quantity discovered from the separability of the Hamilton-Jacobi equation \citep{1968PhRv..174.1559C}. $C$ is relevant to geodesics in the latitudinal direction. We solve $R(r) = dR/dr = 0$ for spherical photon orbits with constant radius $r = \rp$ and obtain two sets of solutions, but only one set of solution is physical and thus relevant to our study \citep{2003GReGr..35.1909T, 2018EPJC...78..879C}:
\begin{subequations}\label{eq8}
\begin{align}
    \Phi &= -\frac{\rp^3 - 3\rp^2 + a^2\rp + a^2}{a(\rp-1)} \label{eq8a} \\
    Q &= -\frac{\rp^3(\rp^3 - 6\rp^2 + 9\rp - 4a^2)}{a^2(\rp-1)^2} \label{eq8b}
\end{align}
\end{subequations}

These parameters are related to the image plane at infinity with orthogonal coordinates $\alpha$ and $\beta$, which observes the KNS at a polar inclination angle $i$ \citep{1972ApJ...178..347B}:
\begin{subequations} \label{eq9}
\begin{align}
    \alpha &= \lim_{r_0\to\infty} \left( -r_0^2 \sin i \frac{d \phi}{dr}\Bigr|_{\substack{r_0, i}} \right) = -\Phi \, \csc i \label{eq9a} \\
    \beta &= \lim_{r_0\to\infty} \left( r_0^2 \frac{d\theta}{dr}\Bigr|_{\substack{r_0, i}} \right) \nonumber\\
    &= \pm \left( Q + a^2 \cos^2 i - \Phi^2 \cot^2 i \right)^{1/2} \label{eq9b}
\end{align}
\end{subequations}

Unstable photon orbits occur when $d^2R/dr^2 < 0$. On the equatorial plane of KBHs, the Carter's constant vanishes and there are two unstable solutions outside the event horizon \citep{1972ApJ...178..347B}:
\begin{align} \label{eq10}
    r_\mathrm{ph\pm} &= 2 + 2 \cos \left[ \frac{2}{3} \arccos (\pm |a|) \right],
\end{align}
where the inner solution $r_\mathrm{ph-}$ and outer solution $r_\mathrm{ph+}$ correspond to the prograde and retrograde equatorial orbit, respectively. For non-equatorial orbits, unstable photon orbits with radii $\rp$ exist in the range $r_\mathrm{ph-} < \rp < r_\mathrm{ph+}$, which distinguish orbits that get captured by the event horizon and those that can return back to infinity. For KNSs, the prograde equatorial orbit does not exist. The retrograde equatorial orbit separates between orbits that terminate at the singularity and those that recede to infinity. Off the equatorial plane, all orbits that approach the KNS can escape. Unstable photon orbits surrounding KNSs exist in the range $\rms < \rp < \rph$, where $\rms$ is the marginally stable radius which satisfies $d^2R/dr^2 = 0$ and $\rph$ is the equatorial retrograde circular radius \citep{2018EPJC...78..879C}:
\begin{subequations}
\begin{align}
    \rms &= 1 + (a^2 - 1)^{1/3} \label{eq11a} \\
    \rph &= 2 + 2 \cosh \left[ \frac{1}{3} \arccosh (2a^2-1) \label{11b} \right]
\end{align}
\end{subequations}

While stable photon orbits with $r < \rms$ are normally hidden inside the event horizon for KBHs, they have physical significance for KNSs due to the lack of an event horizon. Nevertheless, bounded photon orbits cannot be seen by distant observers, so they are irrelevant to our study on the observational features of KNSs.

\subsection{Topological Features of KNS Shadow}
\label{sec2b}

\begin{figure}[t]
      \centering
      \includegraphics[width=0.8\columnwidth,trim=0 24 0 0]{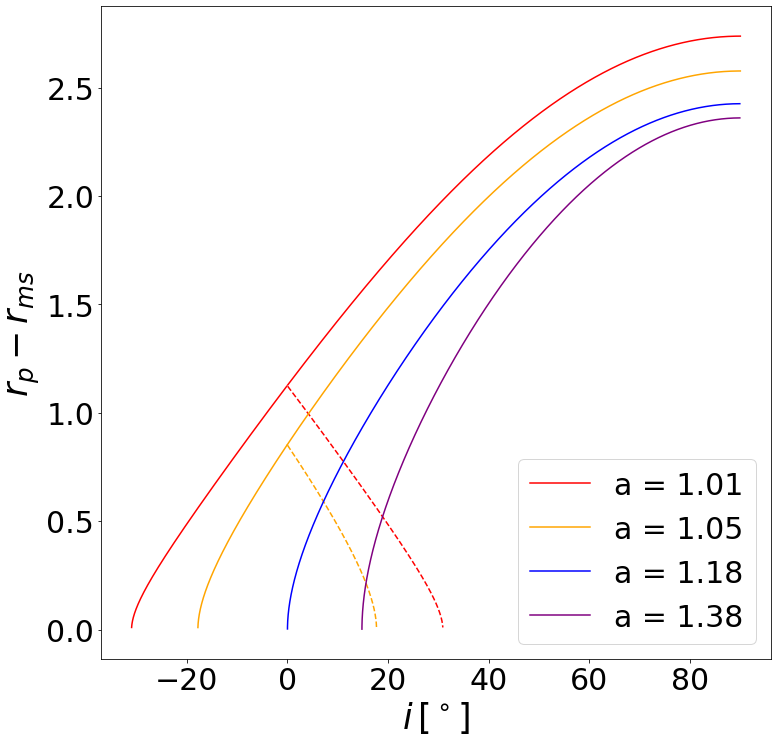}
      \caption{The unstable photon orbit radius $\rp$ (root(s) of $\beta(\rp) = 0$) translated downward by the marginally stable radius $\rms$ for different discrete spins $a = 1.01, 1.05, 1.18,$ and $1.38$ and continuous observational inclination angles $-30^{\circ} \lesssim i < 90^{\circ}$. The solid lines represent unstable root of $\beta(\rp) = 0$. The plot uses axial symmetry $-i \xrightarrow{} i$ to illustrate the second unstable root, denoted in dotted lines, for certain values of $a$ and $i$. The physical implication of the unstable root(s) in the KNS shadow is demonstrated in figure \ref{fig2}}
      \label{fig1}
\end{figure}

The unstable photon orbits with radii $\rms < \rp < \rph$ can be projected to the image plane at infinity using equations~(\ref{eq8}) and~(\ref{eq9}). We define this projection to be the shadow of KNSs. There is a one-to-one correspondence between $\rp$ and the ($\alpha, \beta$) coordinates on the image plane. The shadow of a KBH always has a closed geometry, but the shadow of a KNS might have a gap due to the non-existence of the prograde equatorial circular orbit. The shadow is symmetric with respect to the $\alpha$-axis on the image plane. Thus, we can determine the geometry of the shadow for different values of $a$ and $i$ by solving for $\beta(\rp) = 0$ and count the number of roots in terms of $\rp$, where $\beta$ is expressed as a function of $\rp$ by substituting equations~(\ref{eq8a}) and~(\ref{eq8b}) into equation~\ref{eq9b}, given fixed values of $a$ and $i$. If there are two roots, the shadow is closed. If there is one root, the shadow is open with a gap. If there is no root, the shadow vanishes. Since $\beta(\rp) = 0$ is a sextic polynomial, it does not have an analytical closed-form solution, so we solve this equation numerically and obtain figure~\ref{fig1}.

In figure~\ref{fig1}, we translate the root(s) of $\beta(\rp) = 0$ downward by $\rms$ to illustrate the range of relevant unstable photon orbit radii above the marginally stable radius for different values of $a$ and $i$. As $a$ increases, both $\rp$ and $\rms$ increase; because $\rp$ increases at a slower pace, the $\rp - \rms$ curve shifts downward, so it appears to shift rightward as shown in figure~\ref{fig1}. This shifting pattern of $\rp - \rms$ results in the topological change of the KNS shadow. For $a \lesssim 1.18$, there are unstable roots for a range of $i$ from negative values with small magnitude to $90^{\circ}$, which guarantees the existence of an unstable root. Using the axial symmetry of the Kerr metric, the negative roots can be reflected over the $i = 0^{\circ}$ axis to represent second roots for small $i$ (denoted as dotted lines in figure~\ref{fig1}). This means that for $a \lesssim 1.18$, a shadow can be closed for small $i$ or open for larger $i$. For $a \approx 1.18$, the $\rp - \rms$ curve shifts rightward such that the unstable root of $\beta(\rp) = 0$ only occurs for non-negative $i$, which means that there is only one root, representing an open shadow with a gap. When $a \gtrsim 1.18$, the $\rp - \rms$ curve continues to shift rightward, so there is no unstable root for small $i$, which corresponds to the vanishing of the shadow. As $a$ increases, the minimum $i$ where one unstable root occurs shifts rightward, so the minimum $i$ for the shadow to exist increases. $a \approx 1.18$ is an important critical parameter as it marks the transition between two, one, or zero unstable roots.

\begin{figure}[t]
      \centering
      \includegraphics[width=0.8\columnwidth,trim=0 24 0 0]{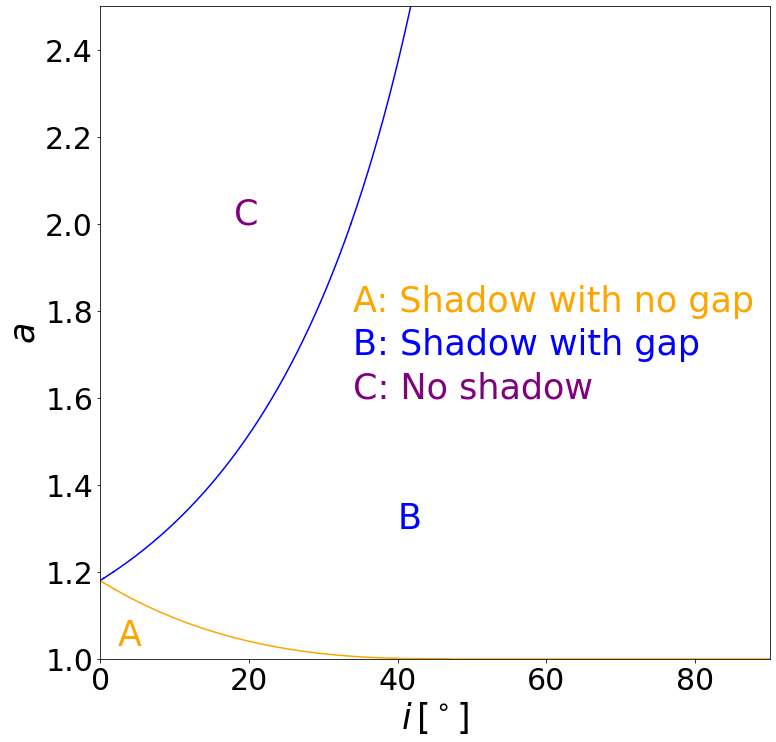}
      \caption{The topological features of KNS shadows for different spins $1 \leq a \leq 2.5$ and observational inclination angles $0^{\circ} \leq i \leq 90^{\circ}$. Regions A, B, and C denote that the shadow is closed, open, or vanishing, respectively. A shadow can only be closed for $a \lesssim 1.18$. The maximum $i$ for a closed shadow (yellow curve) asymptotically approaches $90^{\circ}$ as $a$ decreases towards 1. For $a \gtrsim 1.18$, a shadow cannot be closed and can vanish for certain ranges of $i$. As $a$ increases from $1.18$, the minimum $i$ for the shadow to exist (blue curve) increases.}
      \label{fig2}
\end{figure}

\begin{figure*}
    \centering
    \includegraphics[width=\textwidth]{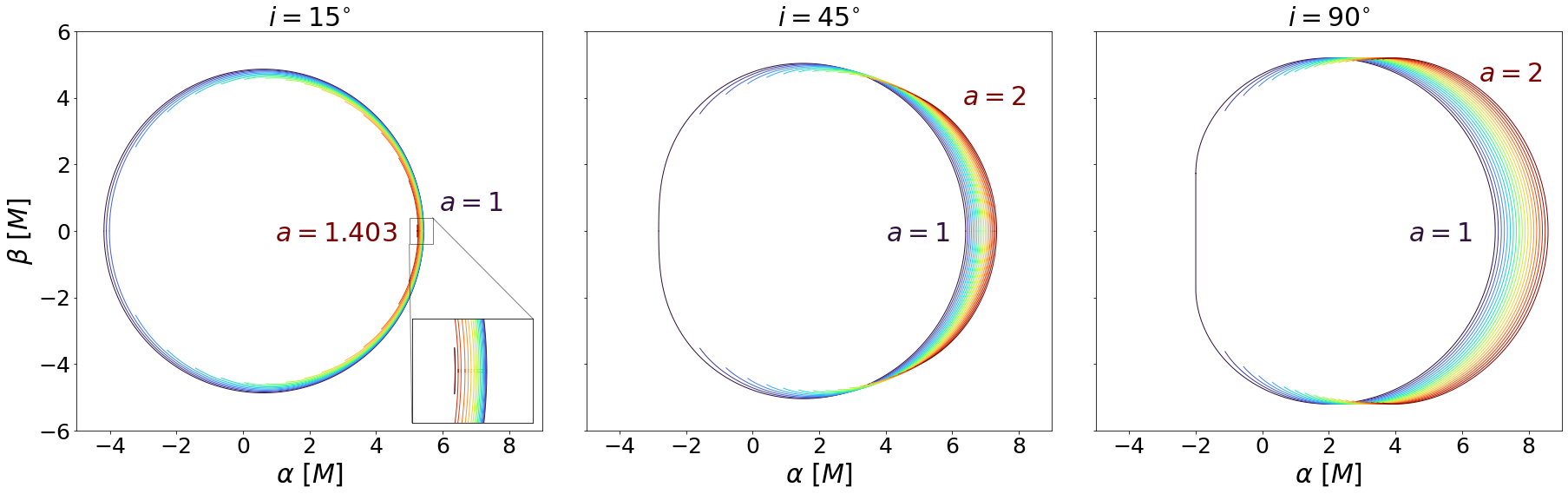}
    \caption{Shadows of KNSs with different spins $a$ and observational inclination angles $i$. From left to right, the plots correspond to $i = 15^{\circ}, \, 45^{\circ}$, and $90^{\circ}$, respectively. In each plot, different colors correspond to different spins, ranging from purple being $a = 1$ to dark red being $a = 1.403$ for $i = 15^{\circ}$ and $a = 2$ for $i = 45^{\circ}$ and $90^{\circ}$. The plots demonstrate that for smaller inclination angles (closer to face-on), the shadow of KNS opens its gap at a greater spin and vanishes at a smaller spin in comparison to greater inclination angles (closer to edge-on). These features are consistent with the $a-i$ phase space classification in figure~\ref{fig2}. In general, the KNS shadows shift rightwards like KBH shadows; however, for low inclination angles like $i = 15^{\circ}$, the shadow vanishes faster than its rightwards shifting, so it appears to shift leftwards.}
    \label{fig3}
\end{figure*}

\begin{figure*}
    \centering
    \includegraphics[width = \textwidth]{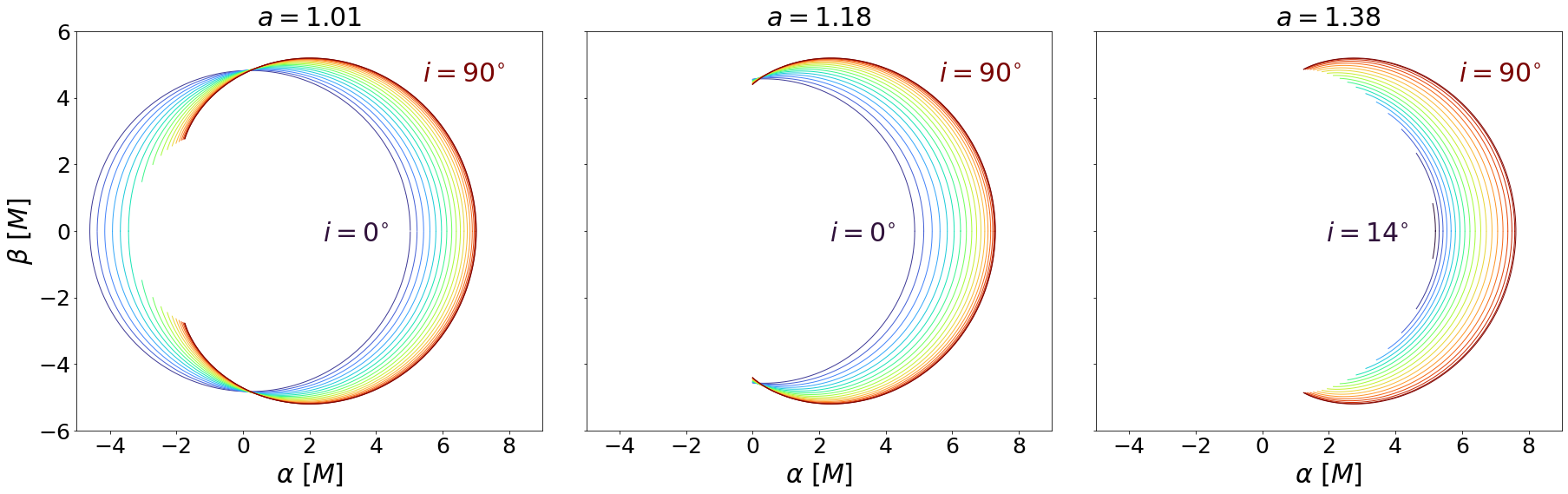}
    \caption{Shadows of KNSs with different spins $a$ and observational inclination angles $i$. From left to right, the plots correspond to $a = 1.01, 1.18$, and $1.38$, respectively. In each plot, different colors correspond to different inclination angles, ranging from purple being $i = 0^{\circ}$ (face-on) for the left and central plot and $i = 14^{\circ}$ for the right plot to dark red being $i = 90^{\circ}$ (edge-on). The plots show that a closed shadow is only possible for lower inclination angles and small spins $1 < a \lesssim 1.18$. For $a \approx 1.18$, the shadow is open with a gap for all inclination angles. For $a \gtrsim 1.18$, the shadow vanishes for small inclination angles. These features are consistent with the $a-i$ phase space classification in figure~\ref{fig2}}
    \label{fig4}
\end{figure*}

The physical implication of the above analysis is summarized in figure~\ref{fig2}, which shows the parameters of $a$ and $i$ where the shadow is closed (region~A), open (region~B), or vanishing (region~C), corresponding to two, one, or zero roots of $\beta(\rp) = 0$, respectively. As $a$ decreases towards $1$ (maximally spinning KBH), the maximum $i$ for a closed shadow asymptotically approaches $90^{\circ}$. This is consistent with the fact that KBHs have closed shadows. For $1 < a \lesssim 1.18$, the shadow can be closed for smaller $i$. As $a$ increases towards $1.18$, the maximum $i$ for a closed shadow decreases to zero. For $a \gtrsim 1.18$, the shadow is no longer closed due to the significant frame dragging (Lense-Thirring) effect for larger spins, and the shadow vanishes for small $i$. As $a$ increases to infinity, the minimum $i$ for the shadow to exist increases. The location ($a \approx 1.18, \, i = 0^{\circ}$) marks the triple point among region A, B, and C on the $a$--$i$ phase space where the topological change in the KNS shadow occurs. Figures~\ref{fig3} and~\ref{fig4} depict the analytical KNS shadow by projecting the unstable photon orbits to an image plane at infinity with orthogonal coordinates $\alpha$ and $\beta$ for different values of $a$ and $i$, according to equations~(\ref{eq8a}) to~(\ref{eq9b}). These figures are consistent with the analytical results in figure~\ref{fig2}.

%==============================================================================
\subsection{Effective Angular Momentum of Unstable Spherical Photon Orbits Around KNS}
\label{sec2c}

\begin{figure}[t]
      \centering
      \includegraphics[width=0.8\columnwidth,trim=0 24 0 0]{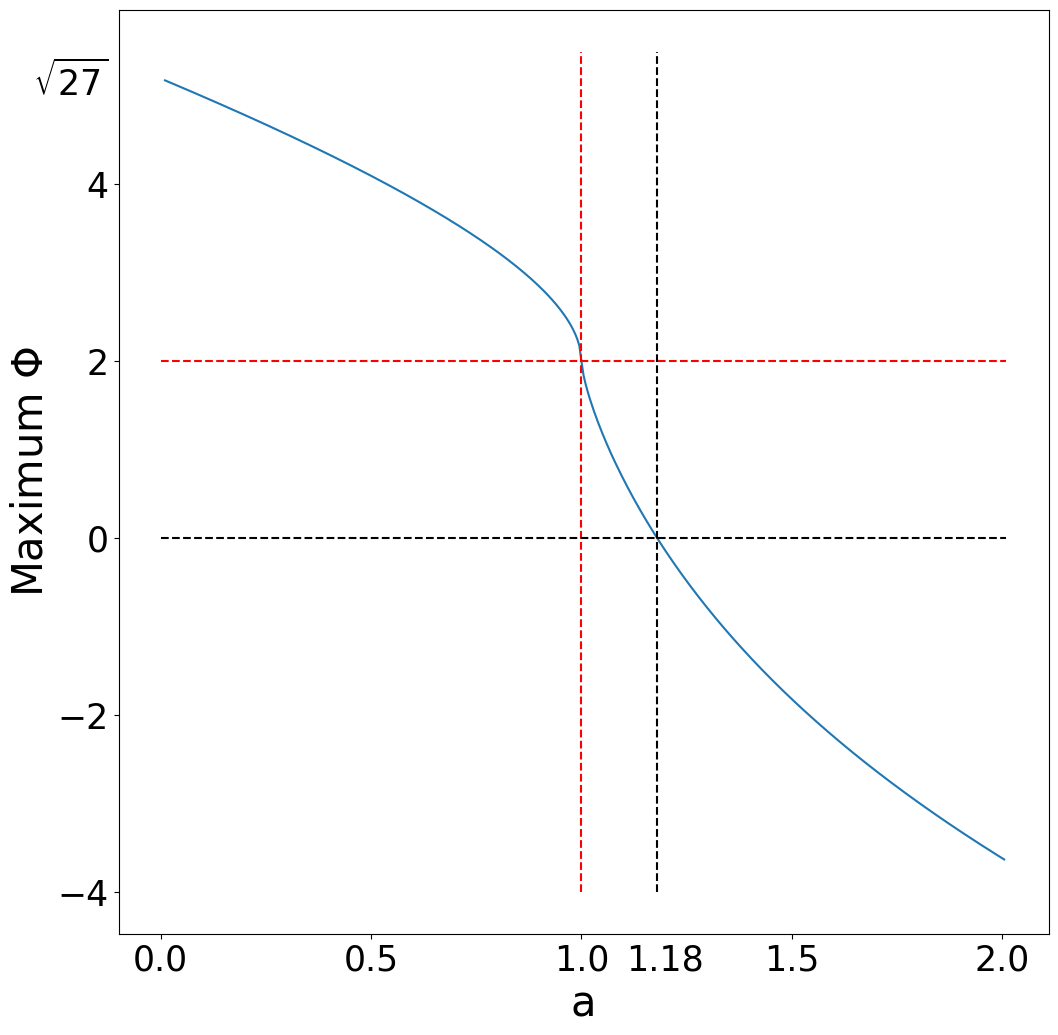}
      \caption{The blue solid line demonstrates maximum $\Phi$ for unstable photon orbits as a function of spin $a$. The red and black dotted lines correspond to spins of interest $a = 1$ and $a \approx 1.18$, respectively.}
      \label{fig5}
\end{figure}

The effective angular momentum of unstable photon orbits $\Phi = L_z/E$, as defined by equation \ref{eq8a} in section \ref{sec2a} can provide further physical insights into the gap opening and vanishing behaviors at different spins and inclinations. For a $<$ 1 (KBH), d$\Phi$/d$r_p$ $<$ 0 outside the event horizon, so the maximum $\Phi$ for unstable photon orbits $\Phi_{max}$ occur at $r_p = r_{ph-}$, the inner equatorial orbit. For a $\geq$ 1 (maximal KBH and KNS), by setting d$\Phi$/d$r_p$ = 0 and d$^2\Phi$/d$r_p^2$ $<$ 0, the $\Phi_{max}$ occur at $r_p = r_{ms}$, the marginally stable orbit. Thus, by substituting $r_p = r_{ph-}$ for $a < 1$ and $r_p = r_{ms}$ for $a \geq 1$ into equation \ref{eq8a}, we can find a function of $\Phi_{max}$ as a function of spin $a$, as plotted in figure~\ref{fig5}.

As the spin increases, the frame dragging effects become stronger so photons that have too much angular momentum would scatter instead of sustaining a spherical orbit around the KBH or KNS. Hence, $\Phi_{max}$ is a decreasing function of spin $a$, which can also be demonstrated mathematically by showing d$\Phi_{max}$/da $<$ 0 in addition to the physical argument above. It is physically significant to examine when $\Phi_{max}$ decreases to 0 where all prograde photon orbits no longer occur. By substituting $r_p = r_{ms}$ into equation \ref{eq8a} and setting $\Phi = 0$: 
\begin{align}\label{eq12}
    a^2 + 3(a^2 - 1)^{2/3} - 3 = 0
\end{align}

\begin{figure*}
    \centering
    \includegraphics[width = \textwidth]{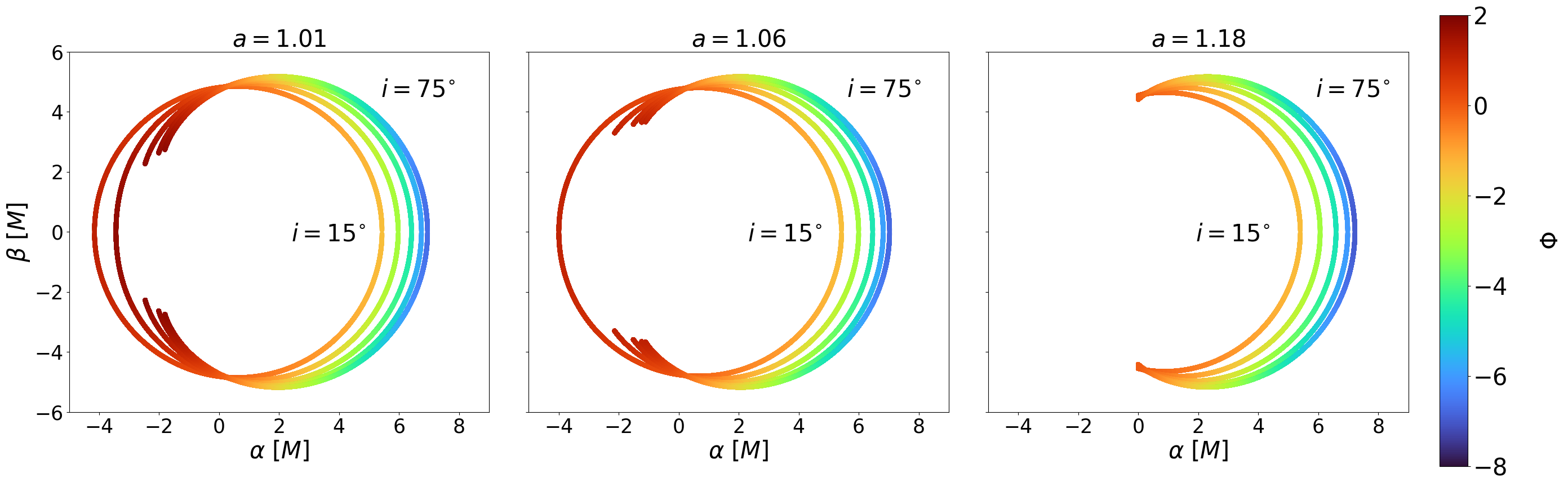}
    \caption{Shadows of KNSs with different spins $a$ and observational inclination angles $i$. From left to right, the plots correspond to $a$ = 1.01, 1.06, and 1.18, respectively. In each plot, the shadow shifts from left to right and ranges from $i = 15^{\circ}$ to $i = 75^{\circ}$, with increments of $15^{\circ}$. The colors represent the effective angular momentum $\Phi$ of each spherical photon orbit projected on the image plane and range from $\Phi = -8$ (retrograde) to $\Phi = 2$ (prograde). Redder and bluer colors represent higher and lower effective angular momentum, respectively. The plots show that for a fixed spin $1 < a \lesssim 1.18$, the shadows can still be closed for lower inclinations due to the existence of prograde spherical photon orbits. As the inclination increases, the prograde photon orbits become more prograde. At some higher inclination angle, their effective angular momentum exceeds $\Phi_{max}$, so they can no longer sustain spherical orbits, and the shadow opens up a gap. At $a \approx 1.18$, all prograde spherical photon orbits no longer exists, so the shadow is always open for all inclinations.}
    \label{fig6}
\end{figure*}

\begin{figure*}
    \centering
    \includegraphics[width = \textwidth]{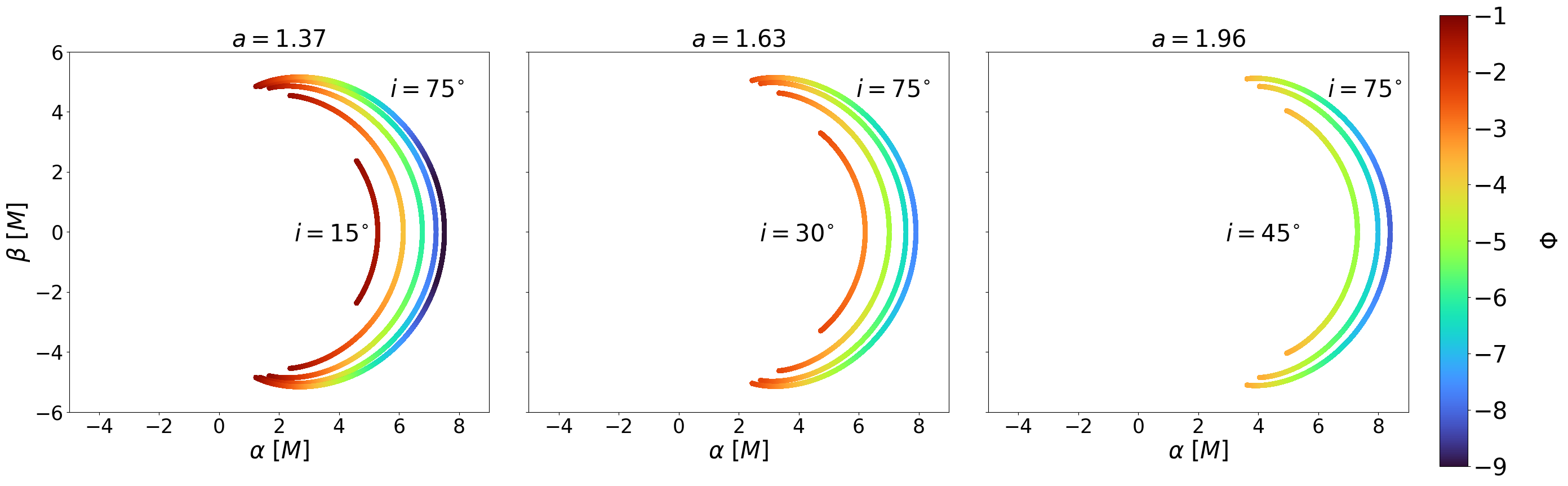}
    \caption{Shadows of KNSs with different spins $a$ and observational inclination angles $i$. From left to right, the plots correspond to $a$ = 1.37, 1.63, and 1.96, respectively. In each plot, the shadow shifts from left to right as the inclination angles range from $i = 15^{\circ}, 30^{\circ}$, and $45^{\circ}$ for the left, center, and right plot, respectively, to $i = 75^{\circ}$, with increments of $15^{\circ}$. The colors represent the effective angular momentum $\Phi$ of each spherical photon orbit projected on the image plane and range from $\Phi = -9$ (more retrograde) to $\Phi = -1$ (less retrograde). Redder and bluer colors represent higher and lower effective angular momentum, respectively. The plots show that for a fixed spin $a \gtrsim 1.18$, as $\Phi_{max} < 0$, retrograde photon orbits at lower inclination might exceed $\Phi_{max}$, so the shadow vanishes. As the inclination increases, the retrograde photon orbits become more retrograde and their effective angular momentum fall below $\Phi_{max}$, so the shadow re-emerges again.}
    \label{fig7}
\end{figure*}

This results in the critical spin $a_{crit} = \sqrt{6 \sqrt{3} - 9} \approx 1.18$, meaning that prograde photon orbits vanish for $a > a_{crit}$. Here, we reproduce the same critical spin where the polar photon orbits disappear as shown in \cite{2018EPJC...78..879C} using a different mathematical approach, and we will take a step further to draw physical connections between photon orbits and KNS shadow geometry. The unstable photon orbits projected to the left and right part of the shadows on the image plane (according to our direction convention) are prograde ($\Phi > 0$) and retrograde ($\Phi < 0$), respectively. The physical picture is that the left side of the shadow consists of prograde photons that travel along the rotational orientation of the KNS, and vice versa. Because $\alpha$ and $\beta$, the orthogonal coordinates of the image plane as defined in equation \ref{eq9a} and \ref{eq9b} in section \ref{sec2a}, are continuous functions of $\Phi$, as $\Phi$ changes from positive to negative values, it smoothly traces out the shadow from left to right. When $a > a_{crit}$, the KNS spins too rapidly such that the prograde photons can no longer sustain spherical orbits, so the left side of the shadow vanishes, and the shadow opens up a gap.

Besides the critical spin, the effective angular momentum of photon orbits can also physically explain the different trends we find in our spin-inclination phase space (figure \ref{fig2}). For a fixed spin, as the inclination increases from face-on to edge-on, the photons have greater (in magnitude) effective angular momentum because it gains motion in the $\phi$ direction. For a fixed spin $1 < a < a_{crit}$, as the inclination increases, the prograde photon trajectories become more prograde. At some higher inclination angle, their effective angular momentum exceeds $\Phi_{max}$ so they can no longer sustain spherical orbits around the KNS. Thus, the shadow opens up its gap at higher inclinations. For a fixed spin $a > a_{crit}$, as $\Phi_{max} < 0$, retrograde photon orbits at lower inclination might exceed the $\Phi_{max}$, so the shadow vanishes. As the inclination increases, the retrograde photon orbits become more retrograde, so they can resist the frame dragging effects and fall below $\Phi_{max}$, allowing them to orbit the KNS and the shadow to re-emerge again. The connections between the effective angular momentum of photons and the projected shadows are visualized in figures \ref{fig6} and \ref{fig7} for different spins and inclinations.

\section{Numerical Experiments}
\label{sec:numerical}

\subsection{Numerical Method}
\label{sec3a}

\begin{figure}
    \centering
    \includegraphics[scale = 0.3]{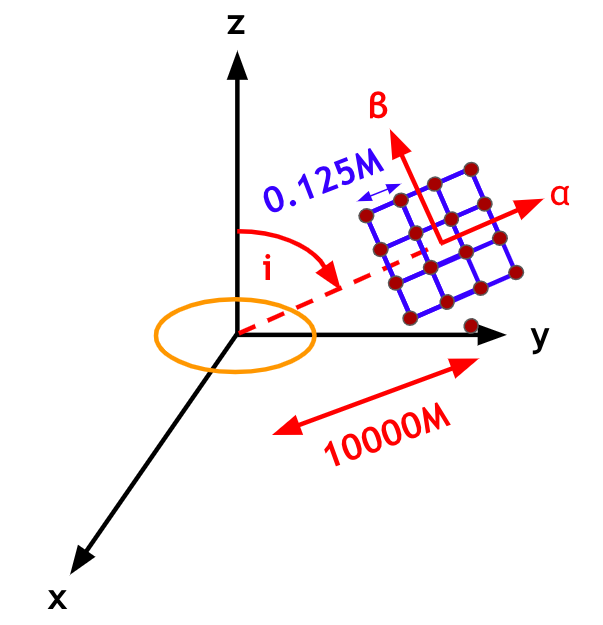}
    \caption{Schematic diagram of the numerical setup. The orange ring represents the KNS centered at the origin of the Cartesian Kerr-Schild coordinate system. The blue square grid represents the image plane, which is located at a distance 10,000$\,M$ away from the KNS at a polar inclination angle $i$ with respect to the normal of the ring singularity. The image plane is effectively at infinity with respect to the KNS, so its orthogonal coordinates $\alpha$ and $\beta$, which is related to the conserved quantities of photons by equations~\ref{eq9}. The brown circles represent the photons, which are spaced $0.125\,M$ from each other on the grid of $128 \times 160$.}
    \label{fig8}
\end{figure}

To confirm our analytical study and obtain better physical insights, we numerically integrate null geodesics in Kerr spacetimes and study their deflection angles.
We place a KNS centered at the origin. The KNS has mass $M$ and dimensionless spin $a$. We set up an image plane located at a distance 10,000$\,M$ away from the KNS at a polar inclination angle $i$ with respect to the $z$-axis, defined to be perpendicular to the ring singularity. The center of the image plane is at the intersection of this plane with the radial vector originating from the singularity. The image plane is effectively at infinity with respect to the KNS, so its orthogonal coordinate $\alpha$ and $\beta$ are related to conserved quantities of the photons---energy $E$, angular momentum $L_z$, and Carter's constant $C$---by equation~(\ref{eq9}). We arrange the photons in a square grid of $128 \times 160$ light rays in the domain $(\alpha, \beta) \in (-8\,M, 8\,M) \times (-8\,M, 10\,M)$ with a spacing of $0.125\,M$. The setup of the image plane is visualized in figure~\ref{fig8}.

We initialize the momentum vectors $\mathbf{k}$ of the photons to be perpendicular to the image plane and satisfy the condition for null geodesics, $k^{\mu}k_{\mu} = 0$ \citep{2013ApJ...777...13C}. We integrate null geodesics backwards in affine parameter starting on the image plane in Cartesian Kerr-Schild coordinates ($t_\mathrm{KS}$, $x$, $y$, $z$), whose Kerr metric is written as \citep{1963PhRvL..11..237K}:
\begin{align}\label{eq13}
    g_{\mu \nu} = \eta_{\mu \nu} + \frac{2Mr^3}{r^4 + a^2z^2} \ell_{\mu} \ell_{\nu},
\end{align}
where $\eta = \mathrm{diag}(-1, 1, 1, 1)$ is the Minkowski metric representing flat spacetime,
\begin{align}\label{eq14}
    \ell_{\mu} = \left( 1, \frac{rx+ay}{r^2 + a^2}, \frac{ry-ax}{r^2 + a^2}, \frac{z}{r} \right),
\end{align}
and the radial component $r$ of the Boyer-Lindquist coordinates can be implicitly defined in Cartesian Kerr-Schild coordinates:
\begin{align}\label{eq15}
    r^4 + (a^2 - R^2) r^2 - a^2z^2 = 0 \mbox{ with } R^2 = x^2 + y^2 + z^2.
\end{align}

We employ the differential geometry software package \texttt{Fadge}~\footnote{\url{https://github.com/adxsrc/fadge}} and the ordinary differential equations solver \texttt{XAJ}~\footnote{\url{https://github.com/adxsrc/xaj}}.
Building on top of Google's GPU-accelerated, composible, and automatic differentiation package \texttt{JAX}~\citep{jax2018github},
\texttt{fadge} automatically derives the geodesic equations from arbitrary metric according to \citet{2018ApJ...867...59C}'s formulation.
\texttt{XAJ} implements the Runge-Kutta Dormand-Prince 4(5) method with an adaptive stepsize control and interpolated dense output~\citep{2002nrca.book.....P}.
Along the photon trajectory, \texttt{XAJ} controls the numerical error by monitoring $k^{\mu} k_{\mu}$ and ensures the photons remain essentially massless for accurate ray tracing calculations.

\begin{figure*}
    \centering
    \includegraphics[width = 0.8\textwidth]{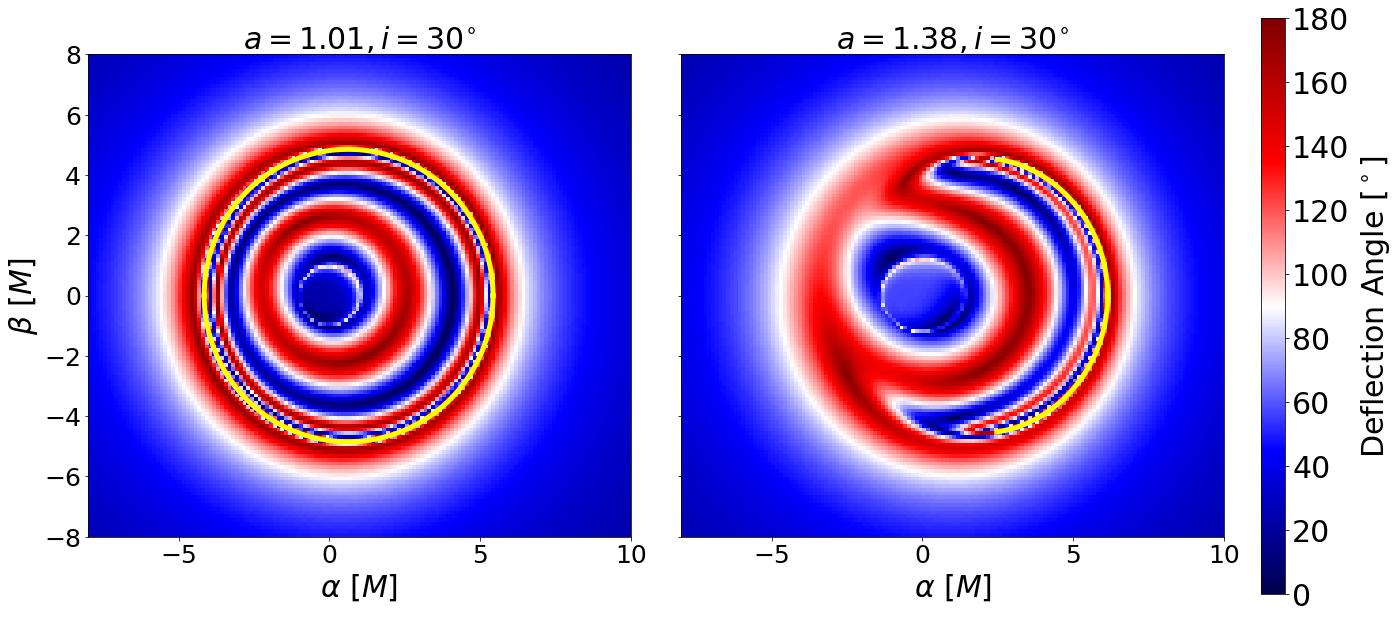}
    \caption{Deflection angle plots on the image plane for spins $a = 1.01$ and $1.38$ and observational inclination angle $i = 30^{\circ}$ for the numerical ray tracing calculations described in section \ref{sec3a}. Redder colors represent higher defelection angles for light rays that return to the vicinity of the image plane. Bluer colors represent lower deflection angles for light rays that travel to the other side of the KNS. The yellow curve represents the analytical shadow computed by equations~(\ref{eq8a})--(\ref{eq9b}) in section~\ref{sec2a}. On the deflection angle plot, the numerical shadow corresponds to the region where light rapidly oscillates between high deflection angles and low deflection angles because the shadow is defined as the projection of unstable spherical photon orbits. Our numerical ray-tracing calculations reproduce the analytical predictions of the KNS shadow in section \ref{sec2b}.}
    \label{fig9}
\end{figure*}

Although it is expected that quantum gravity effects emerge as photons travel very close to a gravitational singularity, our numerical study is purely general relativistic. To avoid any unphysical result that arises from the mathematical limitations of GR and demands correction from a theory of quantum gravity, we terminate the integration if a photon gets too close to the singularity and the numerical instability becomes significant. Because we are interested in the behaviors of unstable spherical photon orbits sufficiently distant from the singularity, this do not affect our analysis of the shadow of KNSs. Also, we do not integrate the radiative transfer equation nor model the emitting plasma around KNSs in this study. We focus on how null geodesics affect the observational features for gravitationally lensed images of KNSs.

For different spins $a$ and observational inclination angles $i$, we compute the deflection angle of every photon on the observer's grid. The deflection angle of a light ray measures the angular difference between the incoming photon that travels towards the KNS and the outgoing photon that travels away from the KNS. For an initial momentum vector $\mathbf{k}_i$ and a final momentum vector $\mathbf{k}_f$, the deflection angle $\theta_d$ is defined by:
\begin{equation}
    \cos \theta_d = \frac{\mathbf{k}_i \cdot \mathbf{k}_f}{\lVert \mathbf{k}_i \rVert \lvert \mathbf{k}_f \rVert}
\end{equation}
From the deflection angle calculations, we compare the numerical ray tracing results with the analytical predictions of the KNS shadow as presented in section~\ref{sec:shadow}.

%==============================================================================
\subsection{Observational Signatures of KNS}
\label{sec3b}

\begin{figure*}
    \centering
    \includegraphics[width = 0.9\textwidth]{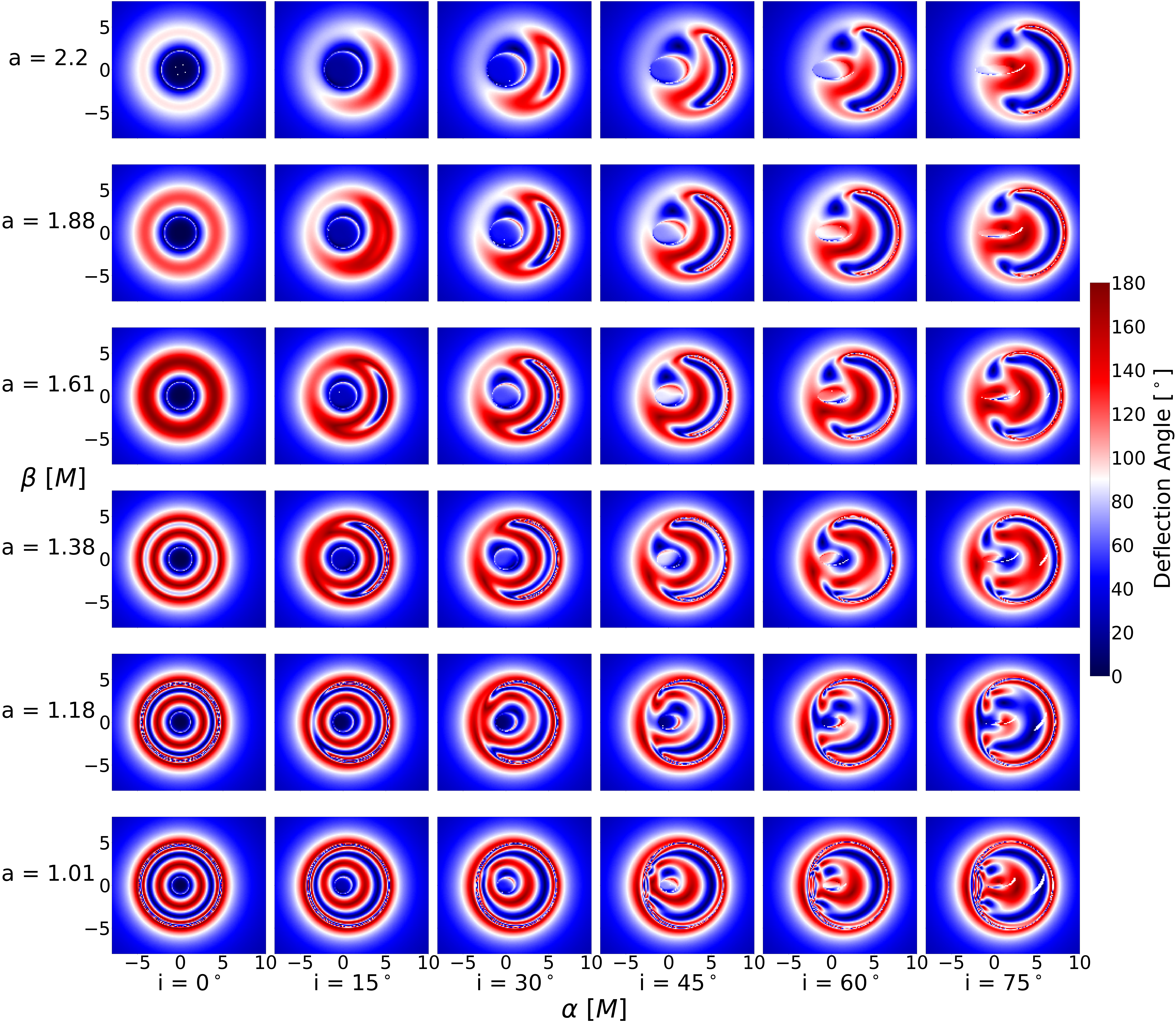}
    \caption{Deflection angle plots on the image plane for different spins $a = 1.01, 1.18, 1.38, 1.61, 1.88$, and $2.2$ (vertical axis) and observational inclination angles $i = 0^{\circ}, 15^{\circ}, 30^{\circ}, 45^{\circ}, 60^{\circ}$, and $75^{\circ}$ (horizontal axis) for the numerical ray tracing calculations described in section \ref{sec3a}. Redder colors represent higher deflection angles for light rays that return to the vicinity of the image plane. Bluer colors represent lower deflection angles for light rays that travel to the other side of the KNS. The spins are approximately chosen according to the relation $a_n = 0.01 + 10^{n/14.7}$, where $n = 0, 1, ..., 5$, so that $a$ is evenly spaced in the logarithmic scale. The plots show that for lower spins $1 < a \lesssim 1.18$, the shadow is closed for lower inclination angles. For $a \approx 1.18$, the shadow opens its gap, and for greater spins $a \gtrsim 1.18$, the shadow vanishes for lower inclination angles. This topological change in the KNS shadow is consistent with the analytical results in figure~\ref{fig2}.}
    \label{fig10}
\end{figure*}

From the numerical ray tracing calculations, we visualize deflection angles of each light ray on the image plane grid for different spins $a$ and observational inclination angles $i$. Figure \ref{fig9} compares between the analytical shadow computed by equations~(\ref{eq8a})--(\ref{eq9b}) in section~\ref{sec2a} and the numerical shadow as an illustration of how to recognize the shadow from the deflection angle visualizations.  Given that the outgoing direction of a photon near an unstable spherical photon orbit is very sensitive to the impact parameter, its deflection angle changes rapidly as function of the impact parameter.  As the shadow is the projection of unstable spherical photon orbits, it corresponds to the region where light rapidly oscillates between getting reflected back to the image plane (high deflection angle) and traveling to the other side of the KNS (low deflection angle). Figure~\ref{fig10} shows the deflection angle plots on the full $a$--$i$ phase space, which reproduces many analytical characteristics of the KNS shadow in section~\ref{sec2b}.

Figure~\ref{fig10} demonstrates that for $1 < a \lesssim 1.18$, the shadow is closed for lower $i$. For the critical parameter $a \approx 1.18$, the shadow opens its gap for all non-zero inclination angles due to the significant frame dragging effect. For $a \gtrsim 1.18$, the shadow vanishes for lower $i$. As the spin increases from $a \approx 1.18$, the minimum $i$ for the shadow to exist increases and the shadow's arc shrinks because frame dragging becomes stronger such that more photons get scattered off instead of orbiting the KNS and returning to the image plane. The topological transition between open, closed, and vanishing of the shadow is consistent with the analytical result shown in figure~\ref{fig2}. Especially, the triple point ($a \approx 1.18, \, i = 0^{\circ}$) is a notable prediction from the analytical calculations in section~\ref{sec2b} reproduced by our numerical ray-tracing calculations. Generally, as $a$ and $i$ increase, the shadow shifts rightwards similarly to KBHs with $a < 1$, except for lower $i$ where the shadow vanishes faster than its rightwards shifting and appears to shift leftwards.

Besides the topological change in the shadow, the absence of an event horizon also results in other distinctive observational features for KNSs. For KBHs, the shadow refers to the apparent boundary that separates between photon orbits that are captured by the event horizon and those that can escape to infinity \citep{2000ApJ...528L..13F}. However, because all photons orbits around KNSs can return to infinity apart for a subset of retrograde equatorial orbits that terminate at the singularity, within the KNS shadow, there are interior structures that behave like mirrors (high deflection angle) where light is deflected back to the image plane and lens (low deflection angle) where the region appears transparent for a distant observer. As $a$ and $i$ increases, these mirror-like and lens-like structure appears to shift continuously. A more thorough study on the topological features of these mirror and lens effect can further provide constraints on observations of SMBHs.

%==============================================================================
\section{Discussions}
\label{sec:discussions}

In this paper, we demonstrate that the shadow of KNSs, defined to be the projection of the unstable spherical photon orbits at infinity, can be closed, open, or vanishing. We analytically study spins $a$ and observational inclination angles $i$ where the shadow possesses these features and changes its topology. We determine that $a \approx 1.18$ is a critical parameter where the KNS shadow can no longer be closed, a distinctive feature that does not happen with the black hole shadow. We further analyze the effective angular momentum of photon orbits to reveal more fundamental physical connections between the light geodesics with the KNS shadow. For $a \gtrsim 1.18$, all prograde orbits can no longer sustain spherical orbits due to the significant frame dragging effects, so the shadow cannot be closed. We also demonstrate that for a fixed spin, as the inclination changes from face-on to edge-on, the photon orbits gain effective angular momentum (in magnitude), which can explain many trends in our spin-inclination phase space (figure \ref{fig2}). For $1 < a \lesssim 1.18$, as the inclination increases, the prograde orbits become more prograde, and their effective angular momentum exceeds the maximum amount $\Phi_{max}$ that can sustain spherical orbits, so the shadow opens up a gap. For $a \gtrsim 1.18$, the shadow might vanish for lower inclinations because retrograde photon orbits might exceed $\Phi_{max} < 0$. As the inclination increases, the retrograde orbits become more retrograde, and their effective angular momentum falls below $\Phi_{max}$, so the shadow re-emerges. Our numerical ray-tracing calculations reproduce these analytical results and provide insights into the observational signatures of KNS images due to gravitational lensing, such as lens-like and mirror-like structures inside the shadow on the image plane.

Because of the linear instability of KNSs in GR \citep{2008CQGra..25s5010C, 2008CQGra..25x5012D, 2018PhLB..780..410N}, our analysis of KNS shadows provide an observational framework to constraint modified gravity theories. If captured by current or future horizon-scale observations of compact object shadows, evidence of KNS signatures can demonstrate violations of one or more assumptions underlying GR and the cosmic censorship conjecture. Considering the likelihood that gravitational singularities are mathematical artifact of GR and the incompatibility between GR and quantum mechanics, predicted observational signatures of KNSs as experimental tests of GR have valuable implications for fundamental physics. Furthermore, our discussion on KNS shadows can be a springboard to study perturbations of KNS spacetime and other types of naked singularities, providing more frameworks to constraint deviations from GR with horizon-scale imaging.
 
While the current EHT images do not have high enough dynamical range to place constraints on KNS for M87* and Sgr~A*, future EHT observations with an enhanced array may. Given that current feature extraction methods from the EHT focuses only on ring-like features \citep{2022PhRvD.106b3017C}, developing image-domain and visibility-domain algorithms to study open rings, therefore, are important for constraining alternative models of KBHs. The $a$--$i$ phase space in figure~\ref{fig2} can be applied to rule out certain ranges of values of $a$ and $i$ depending on whether the shadow is closed, open, or vanishing from shadow-based metric tests of EHT images. Besides, the topological characteristics such as area, curvature radius, and distortion of lens-like and mirror-like regions inside the shadow might provide more thorough measurements and constraints of $a$ and $i$ from observations of Kerr compact objects. The empirical properties of the lens effect is a purely gravitational property of the spacetime independent of the accretion astrophysics around KNSs, so it is highly relevant to astronomical observations. Meanwhile, the observational signatures of the mirror effect also depends on the emitting plasma surrounding KNSs. The lack of surface on the KNS might have repercussions on the behavior of accreting matter and provide unique signatures, such as the repulsive gravitational effect near the singularity demonstrated in \citep{2009PhRvD..80j4023B}. We reserve an exploration of its effects for future studies. From the topological properties of the shadow, the lens, and the mirror structures, we can construct a large, detailed parameter space to compare theoretical predictions of images and light curves of KNSs with future black hole shadow observations to test GR and KNSs as potential candidates of SMBH observations.

%==============================================================================

The authors thank Dimitrios Psaltis, Feryal \"Ozel, Gabriele Bozzola, Dirk Heumann, and Tyler Trent for insightful discussions.
B.N. acknowledges support from an NSF Partnerships for International Research and Education (PIRE) grant OISE-1743747. C.C. acknowledges support from the PIRE grant and an NSF Mid-Scale Innovations Program (MSIP) AST-2034306.

%==============================================================================
\bibliography{refs,main}

%==============================================================================
\end{document}